\documentclass[sigconf]{acmart}

\usepackage{booktabs} 
\usepackage{listings}

\lstdefinestyle{myLuastyle}
{
  language = {[5.2]Lua},
  basicstyle = \ttfamily,
  numbers=left,
  stepnumber=1,
  tabsize=1,
  keywordstyle=\color{magenta},
  stringstyle=\color{blue},
  commentstyle=\color{black!50},
  showstringspaces = false,
  upquote = true,
}
\setcopyright{rightsretained}

\acmDOI{10.475/123_4}

\acmISBN{123-4567-24-567/08/06}

\acmConference[Under review for SIGIR'17]{ACM SIGIR conference}{August 2017}{Tokyo, Japan} 
\acmYear{2017}
\copyrightyear{2017}

\acmPrice{15.00}

\begin{document}
\title{Luandri: a Clean Lua Interface to the Indri Search Engine}

\author{Bhaskar Mitra}
\authornote{The author is a part-time PhD student at UCL.}
\affiliation{%
  \institution{Microsoft, University College London}
  \city{Cambridge, UK}
}
\email{bmitra@microsoft.com}

\author{Fernando Diaz}
\affiliation{%
  \institution{Microsoft}
  \city{New York, USA}
}
\email{fdiaz@microsoft.com}

\author{Nick Craswell}
\affiliation{%
  \institution{Microsoft}
  \city{Bellevue, USA}
}
\email{nickcr@microsoft.com}

\begin{abstract}
In recent years, the information retrieval (IR) community has witnessed the first successful applications of deep neural network models to short-text matching and ad-hoc retrieval. It is exciting to see the research on deep neural networks and IR converge on these tasks of shared interest. However, the two communities have less in common when it comes to the choice of programming languages. Indri, an indexing framework popularly used by the IR community, is written in C++, while Torch, a popular machine learning library for deep learning, is written in the light-weight scripting language Lua. To bridge this gap, we introduce Luandri (pronounced ``\emph{laundry}''), a simple interface for exposing the search capabilities of Indri to Torch models implemented in Lua.
\end{abstract}

%
%



\maketitle


\section{Introduction}
\label{sec:intro}

In recent years, deep neural networks (DNNs) have demonstrated early positive results on a variety of standard information retrieval (IR) tasks, including on short-text matching \cite{huang2013learning, lu2013deep, shen2014latent, hu2014convolutional, severyn2015learning, pang2016text} and ad-hoc retrieval \cite{guo2016deep, mitra2016learning}, and shown promising performances on exciting novel retrieval tasks such as multi-modal retrieval \cite{ma2015multimodal} and conversational IR \cite{yan2016learning, zhou2016multi}. However, the two research communities focused on neural networks and on IR may have less in common when it comes to the choice of programming languages for implementing their respective toolsets. Popular neural network toolkits are often implemented in (or have bindings for) scripting languages, such as Python\footnote{\url{https://www.python.org/}} (e.g., TensorFlow \cite{abadi2016tensorflow}, Theano \cite{bergstra2010theano}, CNTK \cite{yu2014introduction}, Caffe \cite{jia2014caffe}, MXNet \cite{chen2015mxnet}, Chainer \cite{tokui2015chainer}, and PyTorch\footnote{\url{https://github.com/pytorch/pytorch}}) or Lua \cite{ierusalimschy1996lua} (e.g., Torch \cite{collobert2011torch7}) because of their rapid prototyping capabilities. In contrast, many of the popular indexing frameworks for IR are implemented in C++ (e.g., Indri \cite{strohman2005indri}) or Java (e.g., Terrier \cite{ounis2006terrier} and Apache Lucene \cite{mccandless2010lucene}) putting more emphasis on the speed of execution at runtime. The open-source community has developed Python wrappers over the Indri \cite{van2017pyndri} and the Apache Lucene \cite{schreiber2009mixing} programming interfaces to expose the functionalities of these rich IR libraries to the programming language. However, there is still a gap that remains to be bridged for non-Python based deep learning toolkits, such as Torch.

Torch\footnote{\url{https://github.com/torch/torch7}} is a numeric computing framework popular among the deep neural network community. It has been shown to be significantly faster compared to other toolkits such as TensorFlow on convolutional neural networks in multi-GPU environment \cite{shi2016benchmarking}. It is implemented using the light-weight scripting language Lua.\footnote{\url{https://www.lua.org}} In this paper, we introduce Luandri (pronounced ``\emph{laundry}'') -- a Lua wrapper over the Indri search engine. In particular, Luandri exposes parts of the Indri query environment application programming interface (API) for document retrieval including support for the rich Indri query language.

\section{Motivation}
\label{sec:motivation}

\begin{lstlisting}[float=*, xleftmargin=5.0ex, caption={Some Caption}, title={Code snippet 1: Sample Lua code for searching an Indri index using the Luandri API. The Indri index builder application is used for generating the index beforehand. The search query is written using the popular INQUERY structured operators that are supported natively by Indri for specifying matching constraints. The runQuery method in the Luandri API accepts the request as a Lua table and automatically converts it into the appropriate C++ request object that Indri natively expects. Similarly, the result object returned by Indri in C++ is automatically converted to a Lua table.}, label={lst:snippet}]
    local luandri = paths.dofile('luandri.lua')
    local query_environment = QueryEnvironment()
    query_environment:addIndex("path_to_index_file")

    local request = { 
          query = '#syn( #od1(neural networks) #od1(deep learning)) #greater(year 2009)',
          resultsRequested = 10
                    }
    local results = query_environment:runQuery(request).results

    for k, v in pairs(results) do
        print(v.docid .. '\n' .. v.documentName .. '\n' .. v.snippet .. '\n')
    end
\end{lstlisting}

There are a variety of scenarios in which a DNN model can benefit from having access to a search engine during training and/or evaluation. Existing DNN models for ad-hoc retrieval \cite{guo2016deep, mitra2016learning}, for example, operate on query-document pairs to predict relevance. Running these models on the full corpus is prohibitively costly -- therefore the evaluation of these models is often limited to re-ranking top-N candidate documents retrieved by a traditional IR model or a search engine. Typically, these candidate sets are retrieved offline in a process separate from the one in which the DNN is evaluated. However, if the search engine is accessible in the same language as the one in which the DNN is implemented, then the candidate generation step and the DNN-based re-ranking step can follow each other within the same process -- removing the requirement to store large quantity of intermediate datasets containing the candidates to be ranked.

DNN models train on labelled data, although in some cases labels can be inferred rather than explicit. For example, many DNN models for IR \cite{huang2013learning, lu2013deep, shen2014latent, hu2014convolutional, guo2016deep} use negative training examples that are sampled uniformly from the corpus. Recently \citet{mitra2016learning} reported that training with judged negative documents can yield better NDCG performance than training with uniform random negatives. Having access to a search engine during training could enable additional methods for generating negative samples, such as using documents that are retrieved by the engine but at lower ranks.

The lack of adequate labelled data available for training DNN models for ad-hoc retrieval has been a focus for the neural IR community \cite{craswellreport}. It is possible that alternate strategies for supervision may be considered for training these deep models -- including reinforcement learning \cite{williams1992simple} and training under adversarial settings \cite{goodfellow2014generative} -- which could also make use of retrieval from a full corpus during the model training.

\citet{diaz2016query} demonstrated a different application of the traditional retrieval step in the neural IR model. Given a query, they retrieve a set of documents using Indri and use that to train a brand new distributed representation of words specific to that query at run time. Such models, with query-specific representation learning, can be implemented and deployed more easily if the machine learning framework has access to a search engine.

Finally, \citet{ghazvininejad2017knowledge} proposed to ``lookup'' external repositories of facts as part of solving larger tasks using neural network models. Empowering DNN models with access to a search engine may be an exciting area for future exploration.

In all these scenarios, it is useful for a search engine, such as Indri, to be accessible from the same programming language used to implement the DNN. Therefore, we are optimistic that by publicly releasing the Luandri API we will stimulate novel explorations from IR researchers already familiar with Torch.

\section{Querying Indri from Lua}
\label{sec:interface}

Indri is an open-source search engine available with the Lemur toolkit.\footnote{\url{http://www.lemurproject.org/indri/}} Indri consists of two primary components -- an application that builds an index from a raw document collection and another application that can perform searches using this index. The Indri index builder can deal with several different document formats for indexing. This includes TREC (text and Web), HTML, XML, PDF, and plain text among many others.

Searching using Indri involves specifying one or more indices and querying them by either interactively calling the API or by running an application in batch-mode. The Indri query language supports a rich set of operators for specifying phrasal matching conditions, synonymy relationships, document filtering criteria, and other complex constraints. The full query language grammar is available online for reference.\footnote{\url{https://www.lemurproject.org/lemur/IndriQueryLanguage.php}}

Invoking a search on an Indri index using the Luandri API is very similar to how one may use the native C++ Indri API. Code snippet ~\ref{lst:snippet} shows a minimal example of a typical Indri-based search using the Luandri API. We observe that the search is performed by invoking very few lines of Lua code.

The example also demonstrates the use of Indri structured queries. A search is performed using a structured query that constraints the matching to either of the two ordered phrases -- ``neural networks'' or ``deep learning''. The query directs Indri to treat both phrases as synonyms. In addition, a numeric filter is specified to limit matches to only documents whose value corresponding to the year field is greater than 2009.

This example shows searching on the full document index. However, Luandri also allows users to specify a list of document identifiers in the request object to limit the search to only those set of documents. A fixed list of stop words can also be specified for retrieval using the Luandri API. 

The full Luandri implementation is available on GitHub\footnote{\url{https://github.com/bmitra-msft/Luandri}} under the MIT license. We direct interested readers to the source code for exact API specifications.

\section{Under the hood}
\label{sec:method}

The implementation of Lua as a programming language puts a strong emphasis on extensibility \cite{ierusalimschy1996lua}. Lua is an \emph{extension language} because any Lua code can be relatively easily embedded as libraries into code written in other languages. It is also an \emph{extensible language} because of its ability to call functions written in other languages, such as C. The implementation of the Luandri API benefits from the latter property of the language.

Lua comes with a fast Just In Time (JIT) compiler called LuaJIT.\footnote{\url{http://luajit.org}} LuaJIT exposes a foreign-function interface\footnote{\url{http://luajit.org/ext_ffi.html}} (FFI) that makes it easy to call external C functions and manipulate C data structures from Lua. The Luandri API is written using the LuaJIT FFI library.

Luandri API wraps Indri's query environment data types and methods by extern C functions. Then using the LuaJIT's FFI library these C methods are exposed to any code written in Lua. Luandri automatically handles any conversions necessary between Lua tables and Indri's C++ objects, and vice versa. The ``Luandri.cpp'' and ``luandri.lua'' files contain all the wrapper logic on the C++ and the Lua side of our API code, respectively.

The current Luandri API exposes only some of the data structures and methods from Indri's query environment. In future, we hope to expose more of Indri's retrieval functionalities prioritizing based on the need of the broader research community.

\section{Conclusions}
\label{sec:conclusion}

We introduced Luandri, a Lua API to the Indri search engine. Luandri brings to DNN models, implemented on Torch, the retrieval capabilities of Indri, including its powerful query language grammar. We posit that the capabilities of a search engine may be useful for training future DNN models for IR -- for sampling negative examples, or for training under reinforcement or adversarial settings. We hope that the release of Luandri will not only help researchers working on Torch models for IR, but also stimulate new research in novel DNN models that incorporate retrieval from an external knowledge base as an intermediate step towards solving larger tasks.

\bibliographystyle{ACM-Reference-Format}
\bibliography{bibtex} 

\end{document}